%% file: latex/main.tex
\documentclass[11pt]{article}

\usepackage[final]{acl}

\usepackage{times}
\usepackage{latexsym}
\usepackage{amsmath}
\usepackage{placeins}
\usepackage[most]{tcolorbox}
\usepackage{mdframed}
\usepackage{listings}
\tcbuselibrary{breakable}
\usepackage{multirow} 


\usepackage[T1]{fontenc}

\usepackage[utf8]{inputenc}
\DeclareUnicodeCharacter{2191}{\ensuremath{\uparrow}}
\DeclareUnicodeCharacter{2193}{\ensuremath{\downarrow}}

\usepackage{microtype}

\usepackage{inconsolata}

\usepackage{graphicx}
\usepackage{booktabs}
\usepackage{subcaption}

\title{\textsc{ParaStudent}: Closing the Sim2Real Gap in User Simulators for AI Tutor Evaluation}

\author{
  \textbf{Rose Niousha} \quad
  \textbf{Mihran Miroyan} \quad
  \textbf{Abigail O'Neill} \quad
  \textbf{Joseph E. Gonzalez} \\
  \textbf{Gireeja Ranade} \quad
  \textbf{John DeNero} \quad
  \textbf{Narges Norouzi} \\
  University of California, Berkeley \\
  \texttt{\{rose.n, miroyan.mihran, abbyoneill, jegonzal, ranade, denero, norouzi\}@berkeley.edu}
}

\begin{document}
\maketitle
\begin{abstract}
Evaluating Artificial Intelligence (AI) tutor feedback before deployment requires anticipating student engagement, typically assessed through real interaction data. We introduce \textsc{ParaStudent}, a fine-tuning framework for simulating novice programming revisions to support AI tutor evaluation. Compared with prompted baselines, \textsc{ParaStudent}'s revisions more closely match real student code distributions across functional, stylistic, and semantic metrics. Our best variant achieves AUCs of 0.80 for both feedback relevance and successful uptake when distinguishing streams with real engagement above versus at or below the median, while prompted baselines remain near chance on successful uptake. These findings demonstrate the promise of simulated engagement for pre-deployment feedback triage.
\end{abstract}

\input{latex/sections/1_intro}
\input{latex/sections/2_related_work}
\input{latex/sections/3_data}

\input{latex/sections/4_methods}

\input{latex/sections/5_results}
\input{latex/sections/6_discussion}
\input{latex/sections/7_conclusion}
\input{latex/sections/8_limitation}

\bibliography{latex/main}

\input{latex/sections/9_appendix}




\end{document}

%% file: latex/sections/1_intro.tex
\section{Introduction}

\input{latex/figs/contribution}

As Large Language Models (LLMs) are increasingly deployed as interactive assistants, evaluation is shifting beyond isolated turn-level response quality toward measuring how users engage with assistants over time~\cite{zheng2023judging, liao2023rethinking, mysore2025prototypical, li2025beyond}. However, real user interaction data is typically only available after deployment, which can make evaluation slow, expensive, and potentially risky in high-stakes domains.

To address this limitation, recent work has explored LLM-based user simulators for pre-deployment evaluation~\cite{yao2024benchbenchmarktoolagentuser, zhang2025exploring, qian2025userbench}. While such an approach is promising, for simulation-based evaluation to be truly useful, the simulator must reproduce realistic user behavior and support the same evaluation decisions as real interaction data~\cite{suh2026quantifying, zhou2026mind}.

Artificial Intelligence (AI) tutoring is a particularly suitable setting for studying simulation-based evaluation~\cite{yuan2026towards}. Because effective tutoring depends not only on what the tutor says, but also on how students engage with that feedback~\cite{niousha2026missing}, existing AI tutor benchmarks that evaluate tutor responses primarily through static pedagogical qualities such as correctness, style, or solution leakage~\cite{macina2025mathtutorbench,srinivasa2025tutorbench,maurya-etal-2025-unifying} provide only a partial view of tutor effectiveness. Furthermore, pre-deployment evaluation is especially important in education, where deploying ineffective AI tutors not only results in poor user experience but can also directly harm student learning~\cite{jin2025teachtune, ezzaki2025evaluating}.

In particular, AI programming tutors, which are increasingly used by novice learners~\cite{liu2024teaching, kazemitabaar2024codeaid, zamfirescu202561a}, provide a strong setting for simulation-based AI tutor evaluation because learning unfolds through iterative code revisions that provide observable traces of how tutor feedback shapes subsequent student behavior. In addition, prior work shows that simulating programming behavior is more feasible than simulating open-ended dialogue with LLM-based user simulators~\cite{mehri2026measuring}.

We introduce \textsc{ParaStudent} (\autoref{fig:contribution}), a framework for simulating novice programming revisions to evaluate AI tutor feedback before deployment. \textsc{ParaStudent} fine-tunes LLMs conditioned on students' code submissions and AI tutor feedback, with the goal of capturing not only plausible novice code but also how specific feedback shapes subsequent student behavior. We first validate the simulator against real student behavior across functional, stylistic, and semantic dimensions, motivated by prior educational literature, comparing it with existing approaches for simulating novice programmers. Having established this fidelity, we then demonstrate a downstream use case: engagement-based evaluation of AI tutors. Overall, \textsc{ParaStudent} more closely matches real student code than prompted baselines, both distributionally and at the level of paired submissions. Our best variant achieves AUCs of approximately 0.80 for both feedback relevance and successful uptake when distinguishing interaction streams above versus at or below the median real student engagement. These findings demonstrate the promise of simulated student engagement for pre-deployment triage of AI tutor feedback.\footnote{We release our code at \url{https://github.com/rosensh/parastudent}.}

Our contributions are threefold:
\begin{enumerate}
    \item We introduce a fine-tuning method for simulating novice programming revisions conditioned on prior submissions and AI tutor feedback.
    
    \item We propose a multi-dimensional evaluation framework grounded in programming education for validating simulator fidelity across functional, stylistic, and semantic dimensions.
    
    \item We show that distributional fidelity is a prerequisite for meaningful engagement-based evaluation, and that \textsc{ParaStudent}'s simulated engagement scores support pre-deployment triage of AI tutor feedback, with our best variant achieving AUCs of approximately 0.80 for both feedback relevance and successful uptake, substantially above the 0.51--0.58 achieved by prompted baselines for successful uptake.
\end{enumerate}

%% file: latex/figs/contribution.tex
\begin{figure}[t]
    \centering
    \includegraphics[width=\columnwidth]{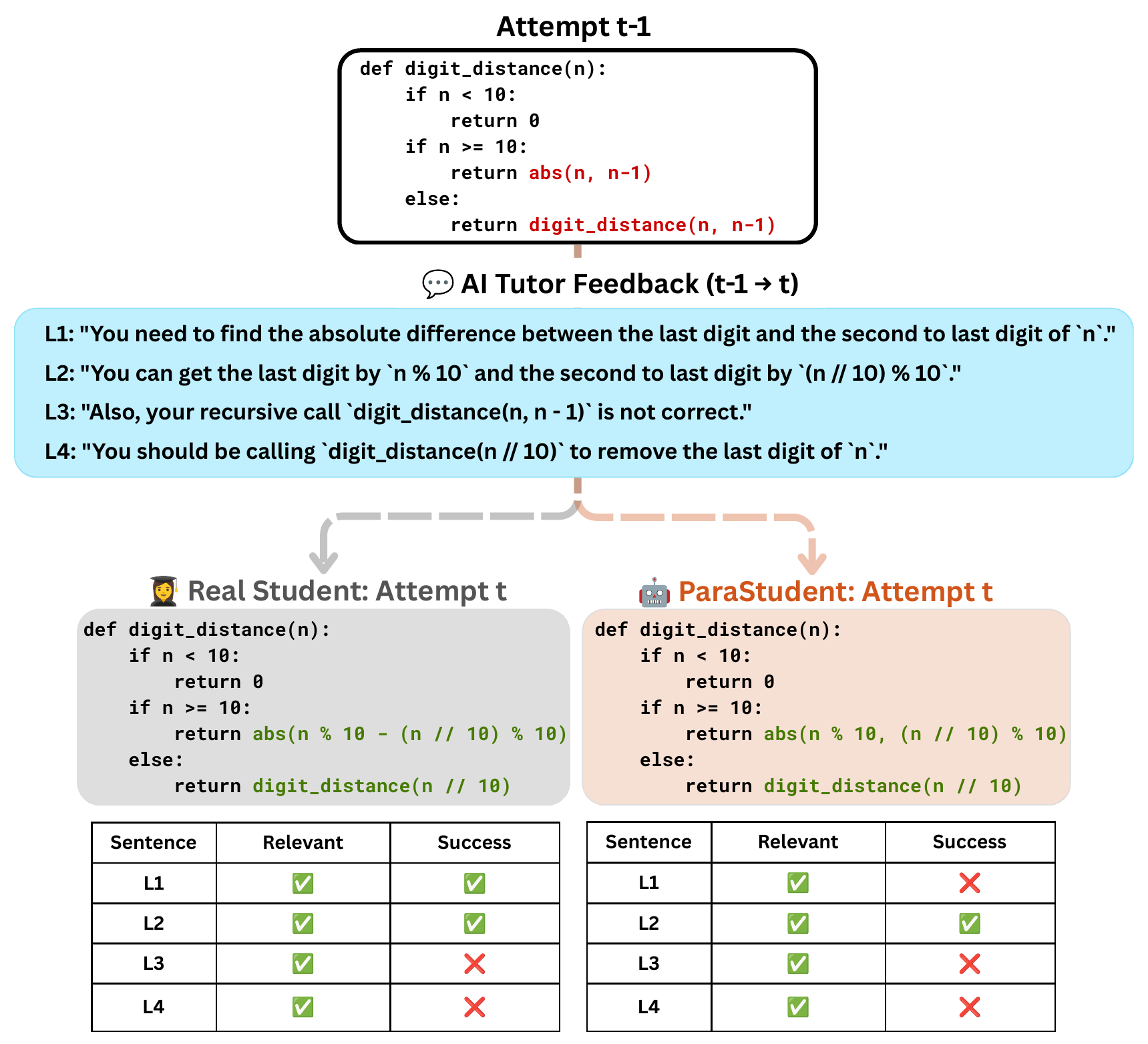}
    \caption{
    \textbf{Overview of real student (left) and \textsc{ParaStudent} (right) edits conditioned on identical prior code and AI tutor feedback.} Each feedback sentence is labeled by whether it influenced the next attempt and whether it was applied correctly; red marks erroneous expressions in attempt \(t{-}1\), green the revised ones in attempt \(t\). Both apply every sentence only partially, and reproducing this incomplete uptake is what enables engagement-based evaluation before deployment.
    }
    \label{fig:contribution}
\end{figure}

%% file: latex/sections/2_related_work.tex
\section{Related Work}
\subsection{LLM-based User Simulation}
\label{sec:related_work_1}

There is a growing body of work using LLMs to simulate human behavior~\cite{piao2025agentsociety,manning2024automated,generativesimulacra,horton2023large,argyle2023out,simulation1000,suh2025language, wang2026simulating}. Existing applications include psychological measurement through interaction with LLM agents~\cite{yang2024psychogat}, large-scale alignment of simulated behavior with expert benchmarks~\cite{wang2025implicit}, and user modeling through inferred beliefs and preferences~\cite{ryan-etal-2025-synthesizeme,singh2025fspo}. Another major application is the training and evaluation of LLM agents, including imagined dialogue with reinforcement learning~\cite{hong2023zero}, proactive and personalized training~\cite{sun2026proactive}, recommender system evaluation~\cite{bougie2025simuser}, and modeling language behavior~\cite{naous2025flipping}.

Despite their popularity, recent work has highlighted the realism gap in LLM-based simulators and argued that evaluation should move beyond surface-level similarity toward distributional validation and downstream utility on real users~\cite{suh2026quantifying,zhou2026mind,mehri2026measuring}. Our work follows this direction by first grounding \textsc{ParaStudent} in real student revision traces, then demonstrating its utility through engagement-fidelity and decision-consistency analyses.

\subsection{Student Simulation}
Building on recent advances, LLM-based simulation has been applied to education in settings such as classroom-level simulation~\cite{classroomsimulacra, simulatelearneractions, zhang2025simulating}, pedagogy and content design~\cite{aguilar2023innovating, li2026type}, pre-service teacher training~\cite{markel2023gpteach, sanyal2025investigating}, using simulated student profiles to evaluate assessment quality~\cite{lu2024generative, sauberli2025llms}, and AI tutor optimization~\cite{dinucu2025problem}. Existing approaches for LLM-based student simulation have primarily relied on prompting~\cite{lu2024generative, sonkar2024llm, student_code_gen2, student_code_gen1}, with more recent work exploring fine-tuning approaches~\cite{ross2025modelingstudentlearning38, ross2025learningmakemistakesmodeling, scarlatos2026simulated}. 

\subsection{AI Tutor Evaluation}
\label{sec:related_work_3}

AI tutoring builds on a long tradition of Intelligent Tutoring Systems (ITS) research. Examples include studies of instructional assistance~\cite{koedinger2007exploring}, tutorial dialogue~\cite{graesser2005autotutor}, programming feedback~\cite{fossati2015data}, and optimization of tutorial tactics~\cite{chi2011evaluation}. This broader literature examines how tutoring supports student learning, beyond the correctness of generated answers.

For LLM-based tutors, recent benchmarks assess pedagogical capabilities~\cite{zhou2025answers,yang2025mmtutorbench,shi2026solver}. MathTutorBench~\cite{macina2025mathtutorbench} and TutorBench~\cite{srinivasa2025tutorbench} evaluate teaching capabilities and core tutoring skills, while MRBench~\cite{maurya-etal-2025-unifying} uses human-annotated dialogues to assess pedagogical quality. TutorGym~\cite{weitekamp2025tutorgym} extends this line to interactive evaluation, but still evaluates tutor actions against predefined tutor policies rather than measuring whether students actually engage with tutor feedback.

We complement these approaches by focusing on students' uptake of specific tutor feedback: whether they ignore it, apply it incorrectly, or incorporate it successfully. This focus follows research emphasizing learners' interpretation and use of feedback~\cite{winstone2017supporting,carless2018development}. Since collecting real student responses involves logistical, ethical, and financial challenges~\cite{yuan2026towards}, we investigate whether \textsc{ParaStudent} can approximate these responses through feedback-conditioned code revision simulations. Our aim is to support pre-deployment triage of AI tutor feedback, complementing pedagogical evaluation and subsequent validation with real students.





%% file: latex/sections/3_data.tex
\section{Data}
We study student code generation in the context of an introductory programming course at the University of California, Berkeley, with an average enrollment of around 900 students per semester. The course covers topics such as functions, recursion, sequences, trees, linked lists, and object-oriented programming. Students complete approximately 10 homework assignments per semester, each with 3--6 problems. Students have access to an autograder system while they complete these assignments that provides immediate feedback on passed and failed test cases with no hidden tests.  Students can keep resubmitting without penalty until their code passes all test cases. As a result, each student–problem pair has a stream of submissions. 

In addition to autograder feedback, our dataset includes feedback from an AI tutor \citep{zamfirescu202561a}, an LLM-based ``Get Help'' assistant deployed in the course to meet common debugging-support needs. The tutor constructs its request from the problem statement, the student's code, and the autograder output. Following typical tutoring patterns \citep{markel2021inside}, its prompt asks the model to assess whether the student understands the problem, which concepts need reinforcement, and whether they have a workable plan, and then to provide conceptual, debugging, or planning support as appropriate. The prompt further instructs the model to keep suggestions short and to withhold the answer and any code. The resulting feedback varies in directiveness, from conceptual scaffolding and goal reminders to concrete error diagnosis and debugging suggestions. Feedback is generated whenever a submission fails at least one test case. All submission attempts are logged, including the student's code, autograder output, and tutor feedback. We focus on assignments implemented in Python.

Our data covers two semesters, Spring 2024 and Spring 2025, restricted to the first five Python homework assignments. We partition by semester: Spring 2025 serves as the training split (445 students, 2,022 submission streams, 6,911 submissions) and Spring 2024 as the test split (479 students, 1,546 submission streams, 6,316 submissions). We hold out the earlier semester because Spring 2024 includes more problems absent from Spring 2025, yielding a larger set of unseen problems for evaluating generalization; the two semesters use comparable assignments and the same deployed AI tutor. \autoref{app:data} gives further detail on IRB compliance and data privacy. 

To evaluate generalization, we define two test subsets:
\textbf{\texttt{test\_OP} (Old Problems)} comprises 5,262 submissions from Spring 2024 students on problems that also appear in the training split, testing generalization to unseen students on familiar problems. \textbf{\texttt{test\_NP} (New Problems)} comprises 1,054 submissions from Spring 2024 students on problems absent from the training data, testing generalization to both unseen students and unseen problems. Alongside aggregate results, we present detailed analyses and visualizations for one representative problem from each subset: \texttt{test\_OP\_1} (1,539 submissions) and \texttt{test\_NP\_1} (695 submissions). Results for two additional problems, \texttt{test\_OP\_2} (968 submissions) and \texttt{test\_NP\_2} (359 submissions), are provided in \autoref{app:results}.





%% file: latex/sections/4_methods.tex
\section{Methodology}
\subsection{Experiments}
\label{sec:exps}
We study the problem of student code generation as follows: given a student $s_i$ (the $i$-th student) and a programming problem $p_u$ (the $u$-th problem), the model must generate the next code submission conditioned on the student’s prior submission history and, when available, the corresponding natural-language feedback instances from an AI tutor. The model sees up to $k$ prior submissions ($k \leq 10$) with or without feedback context, depending on the experiment (we limit the prior submissions to account for the model's context length). Each student–problem pair is represented by a stream of sequential code submissions, from the first to the final attempt.

We evaluate the LLM's ability to generate code under two settings that separate general student-code simulation from feedback-conditioned revision modeling.

\textbf{Experiment 1 (without feedback).}  
Generate the code submission at timestamp $t$ for student $s_i$ on problem $p_u$, conditioned on the student’s prior $k$ code submissions for the same problem:  
\[
c_{t, s_i, p_u} \;\mid\; \big(p_u,\; [c_{t-j, s_i, p_u}]_{j=1...k}\big)
\]

\textbf{Experiment 2 (with feedback).}  
Additionally, condition on the corresponding AI tutor feedback associated with each prior submission:  
\[
c_{t, s_i, p_u} \;\mid\; \big(p_u,\; [c_{t-j, s_i, p_u}, f_{t-j, s_i, p_u}]_{j=1...k}\big)
\]
where $f_{t-j, s_i, p_u}$ denotes the feedback message linked to submission $c_{t-j, s_i, p_u}$.

\subsection{Models}
\label{sec:models}

We compare supervised fine-tuning (SFT) with prompting for student code generation. We use SFT as our primary method because it directly learns from observed student revisions, aligning with our goal of multi-dimensional behavioral imitation. Preference-based methods such as Direct Preference Optimization (DPO)~\cite{rafailov2023direct} offer an alternative, but would require defining preferences that capture the multiple dimensions of student behavior we aim to reproduce. We include prompting baselines since much of prior programming student-simulation work relies on prompting, including the closest work to ours~\cite{student_code_gen2}.

\textbf{Fine-tuning.} We SFT models using LoRA~\cite{lora} ($r=16$, $\alpha=32$), separately for each experiment, for two epochs with a learning rate of $10^{-4}$: Qwen2.5-Coder-7B~\cite{qwen25_coder} for its strong coding capabilities, and Qwen3-8B-Base~\cite{yang2025qwen3} as a complementary general-purpose model. Fine-tuning details are provided in \autoref{app:models}.

\textbf{Prompting.} We evaluate three prompted baselines: instruction-tuned variants of our fine-tuned Qwen models (Qwen2.5-Coder-7B-Instruct and Qwen3-8B), and GPT-5~\cite{gpt5} as a closed-source frontier model. Prompt templates and sampling settings are detailed in \autoref{app:models}.

Throughout the paper, we refer to the five models as follows. The \textsc{ParaStudent} fine-tuned variants are \texttt{qwen-parastudent} (Qwen3-8B-Base) and \texttt{qwen-coder-parastudent} (Qwen2.5-Coder-7B). The prompted baselines are \texttt{gpt-5} (GPT-5), \texttt{qwen-inst} (Qwen3-8B, the instruction-tuned counterpart of Qwen3-8B-Base), and \texttt{qwen-coder-inst} (Qwen2.5-Coder-7B-Instruct). We further conduct both fine-tuning and prompting ablations on Llama-3.1-8B and Llama-3.1-8B-Instruct~\cite{llama3}, with results reported in \autoref{app:results}.

\subsection{Code Fidelity Evaluation}
This section presents a method for evaluating model fidelity by measuring how closely their generated code resembles real student submissions. For each student–problem instance, we generate a single next-attempt sample and evaluate it across multiple metrics grounded in educational evaluation practices.
\label{sec:metrics}

\subsubsection{Functionality Metrics} \label{sec:func} Functional correctness is the primary criterion in many programming education settings, where student submissions are commonly evaluated using autograders. We therefore evaluate each generated submission against the full test suite and compute its pass rate, defined as the fraction of the problem's autograder test cases that the submission passes. To compare simulator fidelity, we report the Wasserstein distance (WD) between real-student and model-generated pass-rate distributions and the Mean Absolute Error (MAE) between paired real and generated pass rates at the same attempt index. Lower values indicate closer alignment for both metrics.

\subsubsection{Style and Code Complexity Metrics}
\label{sec:style}
Programming education also evaluates code beyond correctness, including readability, style, and structural complexity, taught in introductory courses~\cite{de2018understanding, kirk2025distilling}. To capture whether model-generated code reflects these dimensions of student submissions, we extract:

\begin{itemize}
      \item \textbf{PEP 8 violations:} The number of deviations from Python’s style guide (PEP 8)~\cite{pep8}, serving as a proxy for stylistic cleanliness. PEP 8 violations are computed using \texttt{pycodestyle}\footnote{\url{https://pycodestyle.pycqa.org}}.
    \item \textbf{Code length:} Measured by Lines of Code (LOC), reflecting verbosity and conciseness.
    \item \textbf{Abstract Syntax Tree (AST) structure:} AST depth and width~\cite{ast}, which characterize structural complexity.
\end{itemize}

As with pass rate, we report WD between real-student and model-generated distributions and MAE between paired real and generated submissions at the same attempt index for each metric. Lower values indicate closer alignment for both measures.

\input{latex/tables/func-style-agg}

\subsubsection{Embedding Metrics}
\label{sec:emb}
Programming education research has also used code embeddings to capture semantic variation in student solutions and compare patterns across learners~\cite{azcona2019user2code2vec}. We therefore extract 1024-dimensional code embeddings using SFR-Embedding-Code-400M~\cite{embedding_model}, a lightweight model for code retrieval tasks~\cite{coir_leaderboard}.

We compute two embedding-based metrics:
\begin{itemize}
    \item \textbf{Cosine distance:} We compute \(1-\)cosine similarity between each student-written submission and the corresponding model-generated submission for the same attempt, then average across submissions. Lower values indicate closer paired semantic alignment.

    \item \textbf{Top-$k$ coverage:} For each model-generated embedding, we identify its \(k\) nearest student embeddings and report the proportion of unique student samples that appear in at least one such neighborhood. Higher values indicate broader coverage of the student code distribution. We report \(k \in \{5,10\}\).
\end{itemize}

\subsection{AI Tutor Evaluation via Simulated Student Engagement}
\label{sec:utility-method}

After demonstrating its code fidelity, we evaluate whether \textsc{ParaStudent} can support AI tutor feedback evaluation by measuring how simulated students respond to tutor feedback. \citet{niousha2026missing} proposes \textit{engagement scores} as a way to evaluate AI tutors beyond static pedagogical dimensions by grounding evaluation in student interactions. These scores quantify how often the student uses AI tutor feedback and, when used, how accurately it is applied across successive student submissions. We use these scores in two complementary ways. First, we measure \textbf{engagement fidelity} by comparing simulated engagement scores against real student engagement scores using score-level error metrics. Second, we evaluate \textbf{decision consistency}: whether simulator scores discriminate between high- and low-engagement diagnoses derived from real student data. This second analysis reflects a practical pre-deployment tutor-development setting, where instructors may use the simulator to inspect feedback interactions likely to receive weak or successful uptake before collecting real student interaction data.

\paragraph{Engagement score~\cite{niousha2026missing}}
For a given student--problem pair \((s_i, p_u)\), consider the transition from submission \(c_{t-1, s_i, p_u}\) to \(c_{t, s_i, p_u}\). Each tutor feedback message \(f_{t-1, s_i, p_u}\) consists of \(N_t\) sentences \(\{\ell_{t,1}, \ldots, \ell_{t,N_t}\}\), indexed by the transition \(t-1 \rightarrow t\) on which they are evaluated. For each sentence \(\ell_{t,m}\), \(m \in \{1,\ldots,N_t\}\), an LLM judge (GPT-4.1~\cite{gpt41}) assigns a relevance label
\[
\texttt{rel}(c_{t-1,s_i,p_u},\ell_{t,m},c_{t,s_i,p_u})
\in \{0,1\},
\]
indicating whether the sentence influenced the student's subsequent code edit. When \(\texttt{rel}_{t,m}=1\), the judge also assigns a success label
\[
\texttt{succ}(c_{t-1,s_i,p_u},\ell_{t,m},c_{t,s_i,p_u})
\in \{0,1\},
\]
indicating whether the resulting edit successfully applied the feedback. When \(\texttt{rel}_{t,m}=0\), the success label is undefined. We write \(\texttt{rel}_{t,m}\) and \(\texttt{succ}_{t,m}\) for brevity. Sentence-level labels are aggregated into \texttt{RelScore} and \texttt{SuccScore} per feedback message. \texttt{RelScore} measures the proportion of feedback sentences labeled as relevant (\texttt{rel}=1) among all feedback sentences, while \texttt{SuccScore} measures the proportion of successfully applied feedback sentences (\texttt{succ}=1) among the relevant feedback sentences. The judge also produces a natural-language rationale grounded in the code edit before assigning each label. We validate these labels against human annotations on a random sample and find moderate to substantial agreement (see \autoref{app:agreement}).

\input{latex/figs/combined}

Inspired by \citet{dinucu2025problem}, who define tutor rewards over full conversations, we compute engagement at the stream level rather than at the level of individual feedback messages. This reflects the nature of effective feedback, which should support progress over a multi-turn revision trajectory, not merely produce local uptake after a single message~\cite{li2021tracing}. For an interaction stream with \(T\) transitions, we aggregate these message-level scores as:
\[
\bar{S}_{\text{rel}} =
\frac{1}{T}\sum_{t=1}^{T}
\underbrace{\frac{1}{N_t}\sum_{m=1}^{N_t} \texttt{rel}_{t,m}}_{\texttt{RelScore}_t}.
\]

Let \(\mathcal{T}_{\text{rel}}=\{t:\sum_{m=1}^{N_t}\texttt{rel}_{t,m}>0\}\). We define:
\[
\bar{S}_{\text{succ}} =
\frac{1}{|\mathcal{T}_{\text{rel}}|}
\sum_{t \in \mathcal{T}_{\text{rel}}}
\underbrace{
\frac{
\sum_{\substack{1 \leq m \leq N_t\\
\texttt{rel}_{t,m}=1}}
\texttt{succ}_{t,m}
}{
\sum_{m=1}^{N_t}\texttt{rel}_{t,m}
}
}_{\texttt{SuccScore}_t}.
\]
When \(\mathcal{T}_{\mathrm{rel}}\) is empty,
\(\bar{S}_{\mathrm{succ}}\) is undefined, and the stream is excluded from analyses of successful uptake.
Intuitively, \(\bar{S}_{\text{rel}}\) captures how often tutor feedback is taken up in subsequent edits, while \(\bar{S}_{\text{succ}}\) captures how often taken-up feedback is applied correctly across the stream.

\paragraph{Engagement fidelity}
\label{sec:utility-method-a}
Engagement fidelity measures how closely simulated stream-level \(\bar{S}_{\text{rel}}\) and \(\bar{S}_{\text{succ}}\) match the corresponding real student scores. We report MAE and RMSE to capture absolute score error.

\paragraph{Decision consistency}
\label{sec:utility-method-b}
Decision consistency evaluates whether simulated engagement scores distinguish streams with high versus low observed engagement. For each engagement metric, we label streams as high-engagement when their real student score exceeds the $p$-th percentile threshold and low otherwise, with $p \in \{25, 50\}$ to assess sensitivity to the labeling choice. The 25th-percentile threshold distinguishes relatively low-engagement cases from the remaining streams, while the 50th-percentile threshold uses the median. We evaluate continuous simulator scores using ROC-AUC, measuring their ability to discriminate between these real-student engagement labels across simulated-score cutoffs. This assesses discrimination rather than agreement at a single fixed cutoff.

%% file: latex/tables/func-style-agg.tex
\begin{table*}[t]
\caption{\textbf{Fidelity of model-generated code relative to real student submissions on the aggregated test set.} WD is the population-level distance between the real-student and model-generated distributions; MAE is the student-level mean absolute error between each generated submission and the real student's submission at the same attempt index. Lower is better for both. Bold indicates the best WD and the best MAE for each metric within each subset. Across both subsets and every metric, the best value is achieved by a \textsc{ParaStudent} variant, indicating that \textsc{ParaStudent} matches real students both distributionally and at the level of individual submissions. (Experiment 1: without AI tutor feedback; Experiment 2: with AI tutor feedback.)}
\centering

\begin{subtable}[t]{\textwidth}

\centering

\small
\setlength{\tabcolsep}{4.5pt}
\caption{\texttt{test\_OP}}
\label{tab:fidelity-op}
\begin{tabular}{ll cc cc cc cc cc}
\toprule
& & \multicolumn{2}{c}{\textbf{Pass Rate}} & \multicolumn{2}{c}{\textbf{PEP 8}}
& \multicolumn{2}{c}{\textbf{LOC}} & \multicolumn{2}{c}{\textbf{AST Depth}}
& \multicolumn{2}{c}{\textbf{AST Width}} \\
\cmidrule(lr){3-4}\cmidrule(lr){5-6}\cmidrule(lr){7-8}\cmidrule(lr){9-10}\cmidrule(lr){11-12}
\textbf{Model} & \textbf{Exp.}
& WD$\downarrow$ & MAE$\downarrow$ & WD$\downarrow$ & MAE$\downarrow$
& WD$\downarrow$ & MAE$\downarrow$ & WD$\downarrow$ & MAE$\downarrow$
& WD$\downarrow$ & MAE$\downarrow$ \\
\midrule
\texttt{gpt-5}                  & 1 & 0.597 & 0.610 & 1.300 & 1.685 & 1.974 & 3.522 & 1.013 & 1.347 & 3.290 & 5.733 \\
\texttt{qwen-inst}              & 1 & 0.308 & 0.466 & 1.441 & 1.532 & 0.916 & 2.266 & 0.554 & 1.088 & 1.109 & 4.025 \\
\texttt{qwen-coder-inst}        & 1 & 0.334 & 0.482 & 1.156 & 1.429 & 0.716 & 2.188 & 0.337 & 0.986 & 1.083 & 3.940 \\
\texttt{qwen-parastudent}       & 1 & \textbf{0.052} & 0.283 & \textbf{0.125} & 0.640 & \textbf{0.121} & 1.441 & 0.260 & 0.779 & \textbf{0.393} & 2.991 \\
\texttt{qwen-coder-parastudent} & 1 & 0.075 & 0.282 & 0.164 & 0.617 & 0.161 & 1.412 & \textbf{0.139} & 0.770 & 0.414 & 2.827 \\
\midrule
\texttt{gpt-5}                  & 2 & 0.586 & 0.601 & 1.030 & 1.706 & 2.106 & 3.401 & 0.918 & 1.266 & 3.141 & 5.503 \\
\texttt{qwen-inst}              & 2 & 0.336 & 0.465 & 1.391 & 1.513 & 0.531 & 2.122 & 0.492 & 1.019 & 0.828 & 3.845 \\
\texttt{qwen-coder-inst}        & 2 & 0.372 & 0.493 & 1.040 & 1.415 & 0.424 & 2.171 & 0.384 & 0.974 & 0.921 & 3.898 \\
\texttt{qwen-parastudent}       & 2 & 0.080 & \textbf{0.278} & 0.156 & \textbf{0.590} & 0.218 & \textbf{1.353} & 0.175 & \textbf{0.750} & 0.419 & \textbf{2.673} \\
\texttt{qwen-coder-parastudent} & 2 & 0.083 & 0.279 & 0.200 & 0.613 & 0.204 & 1.390 & 0.154 & 0.756 & 0.432 & 2.778 \\
\bottomrule
\end{tabular}
\end{subtable}

\vspace{1em}

\begin{subtable}[t]{\textwidth}
\centering
\small
\setlength{\tabcolsep}{4.5pt}
\caption{\texttt{test\_NP}}
\label{tab:fidelity-np}
\begin{tabular}{ll cc cc cc cc cc}
\toprule
& & \multicolumn{2}{c}{\textbf{Pass Rate}} & \multicolumn{2}{c}{\textbf{PEP 8}}
& \multicolumn{2}{c}{\textbf{LOC}} & \multicolumn{2}{c}{\textbf{AST Depth}}
& \multicolumn{2}{c}{\textbf{AST Width}} \\
\cmidrule(lr){3-4}\cmidrule(lr){5-6}\cmidrule(lr){7-8}\cmidrule(lr){9-10}\cmidrule(lr){11-12}
\textbf{Model} & \textbf{Exp.}
& WD$\downarrow$ & MAE$\downarrow$ & WD$\downarrow$ & MAE$\downarrow$
& WD$\downarrow$ & MAE$\downarrow$ & WD$\downarrow$ & MAE$\downarrow$
& WD$\downarrow$ & MAE$\downarrow$ \\
\midrule
\texttt{gpt-5}                  & 1 & 0.565 & 0.577 & 1.182 & 1.273 & 1.106 & 2.262 & 1.775 & 1.969 & 4.246 & 5.143 \\
\texttt{qwen-inst}              & 1 & 0.406 & 0.524 & 1.009 & 1.261 & 1.962 & 2.171 & 0.403 & 1.151 & 1.659 & 2.925 \\
\texttt{qwen-coder-inst}        & 1 & 0.451 & 0.539 & 0.906 & 1.270 & 0.959 & 2.035 & 0.503 & 1.063 & 0.843 & 3.262 \\
\texttt{qwen-parastudent}       & 1 & 0.100 & 0.302 & 0.107 & 0.511 & \textbf{0.161} & 0.913 & 0.193 & 0.832 & 0.249 & 1.806 \\
\texttt{qwen-coder-parastudent} & 1 & 0.026 & 0.296 & \textbf{0.077} & 0.515 & 0.253 & 0.995 & 0.151 & 0.842 & 0.254 & 1.841 \\
\midrule
\texttt{gpt-5}                  & 2 & 0.562 & 0.576 & 1.166 & 1.246 & 0.283 & 1.973 & 1.090 & 1.379 & 2.309 & 3.588 \\
\texttt{qwen-inst}              & 2 & 0.483 & 0.530 & 0.947 & 1.182 & 1.299 & 1.811 & 0.224 & 1.123 & 0.509 & 2.437 \\
\texttt{qwen-coder-inst}        & 2 & 0.485 & 0.556 & 0.824 & 1.184 & 0.890 & 1.917 & 0.440 & 1.017 & 0.597 & 2.661 \\
\texttt{qwen-parastudent}       & 2 & \textbf{0.007} & 0.286 & 0.089 & \textbf{0.477} & 0.219 & \textbf{0.881} & \textbf{0.146} & 0.779 & \textbf{0.238} & \textbf{1.665} \\
\texttt{qwen-coder-parastudent} & 2 & 0.036 & \textbf{0.270} & 0.120 & 0.494 & 0.384 & 0.959 & 0.176 & \textbf{0.765} & 0.378 & 1.676 \\
\bottomrule
\end{tabular}
\end{subtable}

\label{tab:func-style-agg}
\end{table*}

%% file: latex/figs/combined.tex
\begin{figure*}[t]
\centering
\begin{subfigure}[t]{0.49\textwidth}
    \centering
    \includegraphics[width=\linewidth]{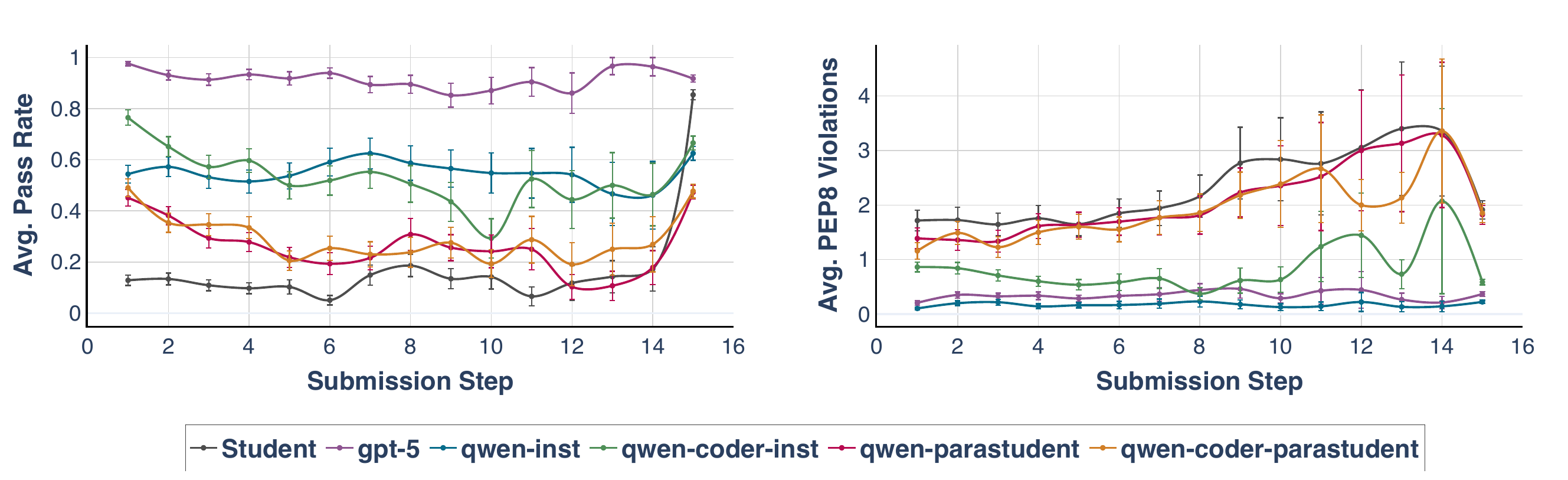}
    \caption{\texttt{test\_OP\_1}}
    \label{fig:combined-a}
\end{subfigure}
\hfill
\begin{subfigure}[t]{0.49\textwidth}
    \centering
    \includegraphics[width=\linewidth]{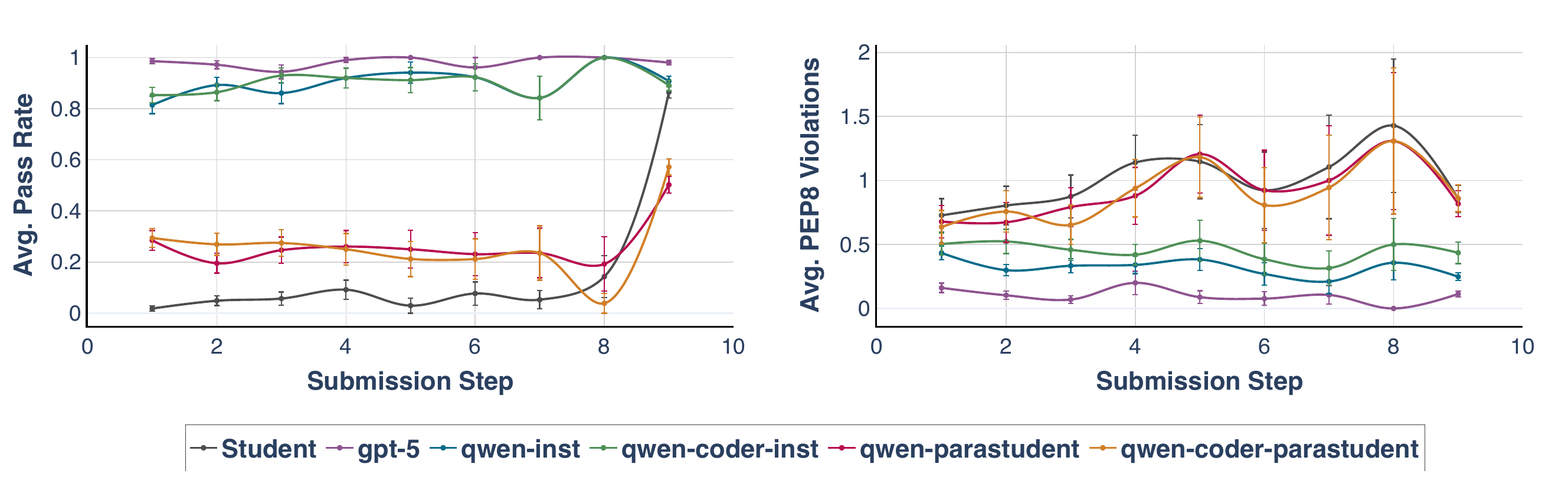}
    \caption{\texttt{test\_NP\_1}}
    \label{fig:combined-b}
\end{subfigure}

\caption{\textbf{Progression of pass rate (left) and PEP 8 violation count (right) across submission steps. (Experiment 2: with AI tutor feedback)} Submission steps correspond to attempt indices; final submissions and submissions at or beyond the 95th-percentile stream-length cutoff are pooled at a shared terminal step. Error bars denote the standard error of the mean over submissions contributing to each plotted point. \textsc{ParaStudent} models most closely align with the student trajectory in both metrics.}
\label{fig:combined}
\end{figure*}

%% file: latex/sections/5_results.tex
\section{Results}

\input{latex/tables/embed}
\input{latex/figs/embed}

\subsection{Code Fidelity}
\label{sec:fidelity}
\subsubsection{Aggregated Results}

We evaluate code fidelity by comparing real and model-generated submissions on the aggregated \texttt{test\_OP} and \texttt{test\_NP} splits using the metrics in \autoref{sec:metrics}.

\textsc{ParaStudent} variants consistently achieve stronger code fidelity than prompted baselines across both previously seen and unseen problems (\autoref{tab:func-style-agg}). Across both splits and experiments, a \textsc{ParaStudent} variant achieves the lowest WD and MAE for every metric. The advantage is particularly pronounced for pass rate, indicating substantially closer agreement with students' functional correctness at both the population and individual-submission levels. Improvements also extend to stylistic and structural metrics.

\subsubsection{Representative Problems}
To illustrate how the aggregate fidelity differences appear within individual problems, we examine submission trajectories and code embeddings for two representative problems drawn from the broader \texttt{test\_OP} and \texttt{test\_NP} subsets: \texttt{test\_OP\_1} (1,539 submissions) and \texttt{test\_NP\_1} (695 submissions), respectively. 

\autoref{fig:combined} shows Experiment~2 submission trajectories for pass rate and PEP~8 violations. Real students begin with low pass rates, improve gradually, and show a sharp increase at the final attempt, a pattern closely matched by both \textsc{ParaStudent} variants. Prompted baselines instead achieve high pass rates from the outset, missing students' iterative trial-and-error behavior. Stylistically, \textsc{ParaStudent} more closely reproduces the persistent PEP~8 violations in student code, whereas prompted baselines produce code with fewer violations.

Embedding-based metrics provide complementary evidence (\autoref{tab:embedding-metrics}). Across both representative problems and experiments, \textsc{ParaStudent} variants achieve lower cosine distances and higher coverage than prompted baselines, indicating closer alignment with real student code in embedding space. On \texttt{test\_OP\_1} (\autoref{tab:embedding-a}), their cosine distances range from 0.014 to 0.017, and top-10 coverage reaches 98.0\%, compared with 71.1\% for the best prompted baseline. The trend is similar on \texttt{test\_NP\_1} (\autoref{tab:embedding-b}). The Experiment~2 projections in \autoref{fig:embd} visually reinforce these findings: \textsc{ParaStudent} representations overlap more closely with real student code across both dense clusters and lower-density regions, whereas prompted baselines exhibit less overlap and narrower coverage.

\subsection{Real vs. Simulated Student Engagement with the AI Tutor}
\label{sec:utility}
We evaluate whether \textsc{ParaStudent} can support pre-deployment assessment of AI tutor feedback on the aggregated \texttt{test\_NP} split, comparing both \textsc{ParaStudent} variants with all three prompted baselines. This held-out new-problem setting approximates evaluating feedback for a new problem before collecting real student interaction data. We focus on Experiment~2, which conditions simulations on the tutor feedback received by students and corresponds to the intended feedback-evaluation setting. 

Following \autoref{sec:utility-method}, we evaluate \textbf{engagement fidelity}, measuring alignment between simulated and real engagement scores, and \textbf{decision consistency}, testing whether simulated scores identify interaction streams with high versus low real engagement. We assess both feedback relevance ($\bar{S}_{\mathrm{rel}}$) and successful uptake ($\bar{S}_{\mathrm{succ}}$). 

\input{latex/figs/engage}
\input{latex/tables/engage_fielity}

\subsubsection{Engagement Fidelity}
\autoref{tab:engagement-fidelity} reports score-level engagement fidelity on the aggregated \texttt{test\_NP} split under Experiment~2. MAE and RMSE quantify deviations between simulated and real stream-level engagement scores.

\texttt{qwen-coder-parastudent} achieves the lowest reported errors for both scores. For relevance $\bar{S}_{\mathrm{rel}}$, it obtains an MAE of 0.13 and RMSE of 0.21, compared with 0.21--0.24 and 0.27--0.30, respectively, for prompted baselines. For successful uptake $\bar{S}_{\mathrm{succ}}$, its MAE is 0.19, and RMSE is 0.31, compared with 0.31--0.32 and 0.42--0.43 for prompted baselines. 

Relative to prompted baselines, both \textsc{ParaStudent} variants achieve lower estimation errors for successful uptake $\bar{S}_{\mathrm{succ}}$, and the coder variant yields the lowest MAE and RMSE across both engagement metrics. These results indicate better agreement with observed engagement, suggesting an advantage over prompted models in approximating how students engage with tutor feedback.

\subsubsection{Decision Consistency}
\autoref{fig:engage} shows ROC curves and AUCs on the aggregated \texttt{test\_NP} split under Experiment~2. We evaluate whether simulated scores distinguish interaction streams whose real engagement exceeds the 25th- or 50th-percentile threshold. Corresponding 95\% confidence intervals are reported in \autoref{app:results}.

The clearest separation between \textsc{ParaStudent} and prompted baselines occurs for successful uptake $\bar{S}_{\mathrm{succ}}$. At both thresholds, \texttt{qwen-coder-parastudent} achieves an AUC of approximately 0.80 and \texttt{qwen-parastudent} achieves approximately 0.72, compared with 0.51--0.58 across prompted baselines. For relevance $\bar{S}_{\mathrm{rel}}$, differences are smaller: the coder variant achieves AUCs of 0.81 and 0.80 at the 25th- and 50th-percentile thresholds, respectively, compared with 0.74--0.75 and 0.76--0.77 for prompted baselines.

This pattern is consistent with the code-fidelity results (\autoref{sec:fidelity}), where prompted baselines produce overly successful revisions and miss students' gradual trial-and-error behavior. Under Experiment~2, all models receive AI tutor feedback and can generate edits related to it; relevance alone therefore does not establish that a simulator reproduces novice feedback uptake. Successful uptake additionally measures whether taken-up feedback is applied correctly. The stronger discrimination of \textsc{ParaStudent} for $\bar{S}_{\mathrm{succ}}$ is thus consistent with better modeling of students' imperfect application of feedback. 


Practically, these findings suggest a role for \textsc{ParaStudent} in pre-deployment triage of AI tutor feedback. Simulated engagement scores can help prioritize feedback for instructor review, while generated revision trajectories and sentence-level uptake labels illustrate how students might respond. Instructors can inspect simulated cases of feedback being ignored, applied incorrectly, or successfully incorporated, and use these examples to evaluate and refine AI tutor feedback. 

%% file: latex/tables/embed.tex
\begin{table*}[t]
\caption{\textbf{Embedding-based Similarity Metrics.} Cosine distance measures per-submission alignment with the real student code, and top-$k$ coverage indicates the proportion of student codes represented in model neighborhoods (see \autoref{sec:emb} for details). Lower distances and higher coverage indicate better alignment with student code distribution. The \textsc{ParaStudent} variants consistently align most closely with the real student distribution across all metrics. Bold indicates the best result for each metric across both experiments. (Experiment 1: without AI tutor feedback, Experiment 2: with AI tutor feedback).}
\label{tab:embedding-metrics}
\centering
\scriptsize
\setlength{\tabcolsep}{4pt}
\renewcommand{\arraystretch}{0.85}
\begin{subtable}[t]{0.48\textwidth}
\caption{\texttt{test\_OP\_1}}
\label{tab:embedding-a}
\centering
\begin{tabular}{l|c|c|cc}
\toprule
\textbf{Model} & \textbf{Exp.} & \textbf{Cosine Distance ↓} & \multicolumn{2}{c}{\textbf{Coverage (\%) ↑}} \\
 & & & \textbf{k=5} & \textbf{k=10} \\
\midrule
\texttt{gpt-5}                  & 1 & 0.035 & 20.5 & 28.7 \\
\texttt{qwen-inst}              & 1 & 0.028 & 30.4 & 46.0 \\
\texttt{qwen-coder-inst}        & 1 & 0.028 & 56.7 & 71.1 \\
\texttt{qwen-parastudent}       & 1 & 0.016 & 93.7 & 97.4 \\
\texttt{qwen-coder-parastudent} & 1 & 0.016 & \textbf{94.0} & \textbf{98.0} \\
\midrule
\texttt{gpt-5}                  & 2 & 0.029 & 51.1 & 66.8 \\
\texttt{qwen-inst}              & 2 & 0.022 & 53.9 & 70.0 \\
\texttt{qwen-coder-inst}        & 2 & 0.027 & 55.9 & 70.8 \\
\texttt{qwen-parastudent}       & 2 & \textbf{0.014} & 92.8 & 97.5 \\
\texttt{qwen-coder-parastudent} & 2 & 0.017 & 92.8 & 97.3 \\
\bottomrule
\end{tabular}
\end{subtable}
\hfill
\begin{subtable}[t]{0.48\textwidth}
\caption{\texttt{test\_NP\_1}}
\label{tab:embedding-b}
\centering
\begin{tabular}{l|c|c|cc}
\toprule
\textbf{Model} & \textbf{Exp} & \textbf{Cosine Distance ↓} & \multicolumn{2}{c}{\textbf{Coverage (\%) ↑}} \\
 & & & \textbf{k=5} & \textbf{k=10} \\
\midrule
\texttt{gpt-5}                  & 1 & 0.053 & 29.5 & 45.6 \\
\texttt{qwen-inst}              & 1 & 0.045 & 36.7 & 54.8 \\
\texttt{qwen-coder-inst}        & 1 & 0.042 & 60.7 & 74.8 \\
\texttt{qwen-parastudent}       & 1 & 0.021 & \textbf{94.7} & \textbf{98.1} \\
\texttt{qwen-coder-parastudent} & 1 & 0.022 & 94.1 & \textbf{98.1} \\
\midrule
\texttt{gpt-5}                  & 2 & 0.043 & 44.5 & 63.7 \\
\texttt{qwen-inst}              & 2 & 0.039 & 46.6 & 64.8 \\
\texttt{qwen-coder-inst}        & 2 & 0.038 & 60.7 & 75.7 \\
\texttt{qwen-parastudent}       & 2 & \textbf{0.018} & 91.4 & 96.1 \\
\texttt{qwen-coder-parastudent} & 2 & 0.021 & 92.5 & 97.6 \\
\bottomrule
\end{tabular}
\end{subtable}
\end{table*}

%% file: latex/figs/embed.tex
\begin{figure*}[t]
    \centering
    \begin{subfigure}[t]{0.49\textwidth}
        \centering
        \includegraphics[width=\linewidth]{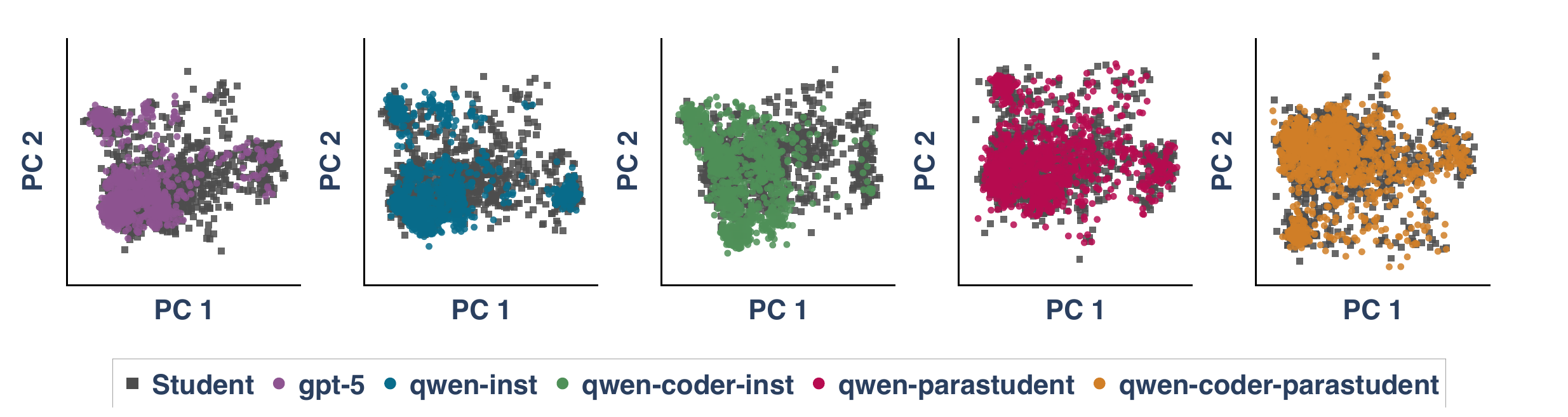}
        \caption{\texttt{test\_OP\_1}}
        \label{fig:embd-a}
    \end{subfigure}
    \hfill
    \begin{subfigure}[t]{0.49\textwidth}
        \centering 
        \includegraphics[width=\linewidth]{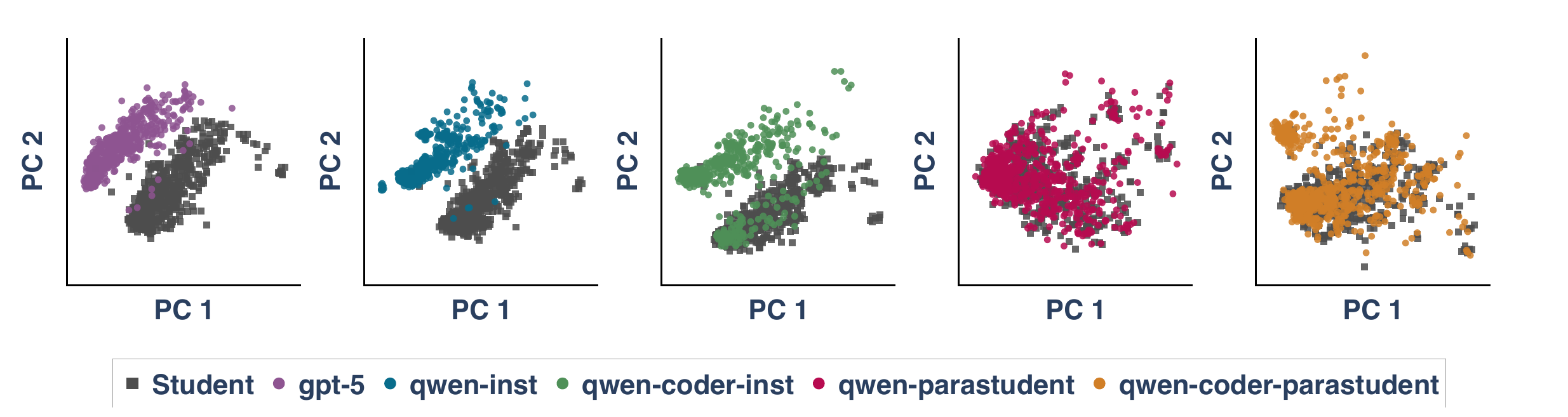}
        \caption{\texttt{test\_NP\_1}}
        \label{fig:embd-b}
    \end{subfigure}

\caption{\textbf{Code embedding visualization (Experiment 2: with AI tutor feedback)}. Real student code (gray squares) is compared against \texttt{gpt-5} (purple circles), \texttt{qwen-inst} (blue circles), \texttt{qwen-coder-inst} (green circles),  \texttt{qwen-parastudent} (pink circles),  and \texttt{qwen-coder-parastudent} (orange circles). 1024-dimensional embeddings are projected onto a 2D plane using Principal Component Analysis (PCA) for visualization purposes. \textsc{ParaStudent} variants show the substantially stronger overlap with the student code distribution compared to baselines, including better coverage of outlier codes where there is less clustering.}
    \label{fig:embd}
\end{figure*}

%% file: latex/figs/engage.tex
\begin{figure*}[t]
\centering
\includegraphics[width=\textwidth]{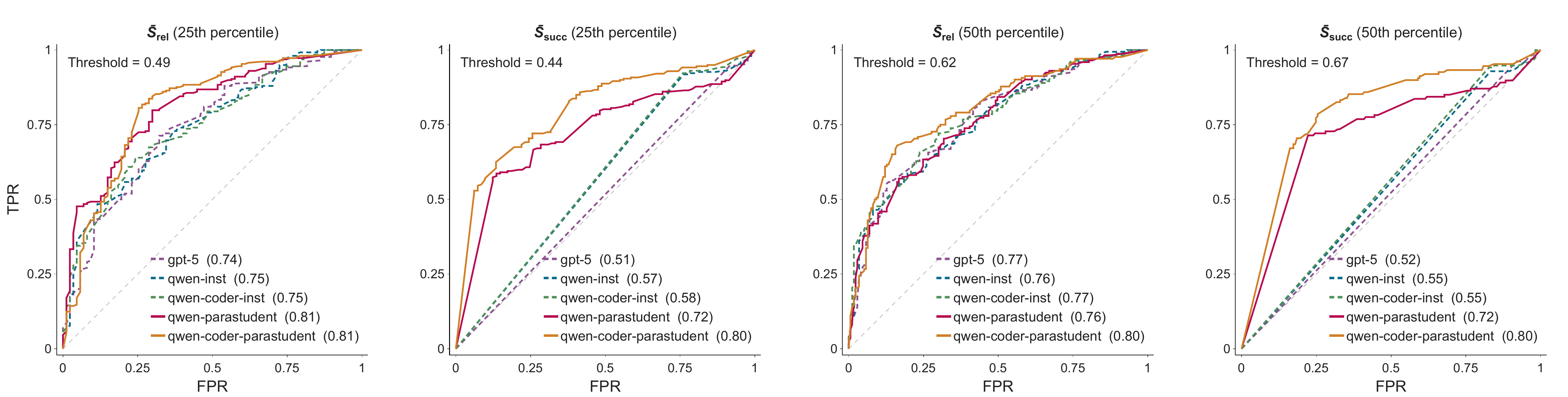}
\caption{\textbf{Engagement ROC curves on the aggregated \texttt{test\_NP} split.} All models use tutor feedback (Experiment~2). From left to right, panels show relevance $\bar{S}_{\mathrm{rel}}$ and successful uptake $\bar{S}_{\mathrm{succ}}$ at the 25th-percentile threshold, followed by the same metrics at the 50th-percentile threshold. Thresholds are computed from real engagement scores; panel annotations give threshold values, and legends report AUCs. Dashed lines denote prompted baselines; solid lines denote \textsc{ParaStudent} variants. Both variants achieve higher AUCs for successful uptake at both thresholds.}
\label{fig:engage}
\end{figure*}

%% file: latex/tables/engage_fielity.tex
\begin{table}[t]
\centering
\caption{\textbf{Simulated engagement fidelity on the aggregated \texttt{test\_NP} split.} All models use Experiment~2 (with AI tutor feedback). Lower is better.}
\label{tab:engagement-fidelity}
\small
\setlength{\tabcolsep}{4pt}
\begin{tabular}{@{}c|l|c|c@{}}
\toprule
\textbf{Score} & \multicolumn{1}{c|}{\textbf{Model}}& \textbf{MAE $\downarrow$} & \textbf{RMSE $\downarrow$} \\
\midrule
$\bar{S}_{\mathrm{rel}}$ & \texttt{gpt-5} & 0.21 & 0.27 \\
$\bar{S}_{\mathrm{rel}}$ & \texttt{qwen-inst} & 0.24 & 0.30 \\
$\bar{S}_{\mathrm{rel}}$ & \texttt{qwen-coder-inst} & 0.23 & 0.29 \\
$\bar{S}_{\mathrm{rel}}$ & \texttt{qwen-parastudent} & 0.22 & 0.28 \\
$\bar{S}_{\mathrm{rel}}$ & \texttt{qwen-coder-parastudent} & \textbf{0.13} & \textbf{0.21} \\
\midrule
$\bar{S}_{\mathrm{succ}}$ & \texttt{gpt-5} & 0.32 & 0.43 \\
$\bar{S}_{\mathrm{succ}}$ & \texttt{qwen-inst} & 0.32 & 0.43 \\
$\bar{S}_{\mathrm{succ}}$ & \texttt{qwen-coder-inst} & 0.31 & 0.42 \\
$\bar{S}_{\mathrm{succ}}$ & \texttt{qwen-parastudent} & 0.24 & 0.37 \\
$\bar{S}_{\mathrm{succ}}$ & \texttt{qwen-coder-parastudent} & \textbf{0.19} & \textbf{0.31} \\
\bottomrule
\end{tabular}
\end{table}

%% file: latex/sections/7_conclusion.tex
\section{Conclusion}
We introduced \textsc{ParaStudent} for pre-deployment AI tutor evaluation. By fine-tuning on real revision trajectories, it matches real student code more closely than prompted baselines across functional, stylistic, and semantic dimensions, and its simulated engagement scores better discriminate high- versus low-uptake streams. Together, these findings show how simulated engagement can support pre-deployment triage of AI tutor feedback.

%% file: latex/sections/8_limitation.tex
\section*{Limitations}
While our findings highlight the importance of fine-tuning and holistic evaluation of LLMs in student simulation, several limitations must be acknowledged.

\begin{itemize}
    \item This work is limited to one course (with a large enrollment of around 900 students). The training and test datasets are drawn from different semesters of the same course. Obtaining a comparable longitudinal dataset with iterative student submission and associated AI tutor feedback is particularly challenging in the education domain. 
    We note, however, that our \texttt{test\_NP} setting — where the simulator is evaluated on problems unseen during training — provides a partial signal of generalizability, since it requires the model to simulate student behavior on novel problem structures without problem-specific fine-tuning. Extending evaluation to other courses, institutions, and programming languages is an important next step.
    \item A more comprehensive comparison across a broader range of model types, sizes, and families is an important direction for future research. Nevertheless, these baselines are representative of commonly used instruction-tuned and frontier models, and the consistent trends observed across them support our conclusions about the limitations of prompt-based student simulation. The consistent advantage of fine-tuning over prompting across both model families tested (Qwen and Llama) suggests this finding is unlikely to be model-specific, though verification on additional architectures remains open.
    
\end{itemize}

\paragraph{Potential Risks.} While simulating student code can offer pedagogical benefits, it also raises several risks. First, if misapplied, such models could reinforce incorrect programming habits or misconceptions by overfitting to common student errors. Additionally, any deployment of such systems in educational settings must be done with care, including appropriate safeguards for data anonymization and ethical use.

%% file: latex/sections/9_appendix.tex
\appendix

\input{latex/figs/ft-prompt}

\input{latex/figs/prompting-prompt}
\section{Data}
\label{app:data}

\textbf{Privacy.} We received IRB approval for the research use of the dataset (protocol ID: 2023-09-16725), after which the AI tutor was deployed in the course. When students submitted their code to the autograder as part of the regular course practice, they were asked whether they would like to receive AI tutor feedback on their code, and if so, were asked whether they would like their data to be used for research. Students were still able to receive AI feedback even if they did not wish their data to be used in research. Only students who explicitly consented were included in our analysis. Additionally, all student identifiers (e.g., IDs and emails) were fully anonymized prior to passing the data to the LLM and any further analysis. As part of our system design, any data from LLM calls from our educational platform to the model providers will not be retained on third-party servers and will not be used for further fine-tuning of those models.

\textbf{Documentation.} Data contains student code submissions for an introductory programming course from the University of California, Berkeley. We filter data from the Spring 2024 and Spring 2025 semesters. Additionally, we filter only assignment problems in Python. Student demographic information is not available due to privacy regulations. 

\section{Models}
\label{app:models}

\textbf{Models.} We used the following models for fine-tuning experiments: Qwen2.5-Coder-7B \cite{qwen25_coder}, Qwen3-8B-Base~\cite{yang2025qwen3}, and Llama-3.1-8B~\cite{llama3}. We used the following models for prompting experiments: Qwen2.5-Coder-7B-Instruct \cite{qwen25_coder}, Qwen3-8B~\cite{yang2025qwen3}, GPT-5 \cite{gpt5}, and Llama-3.1-8B-Instruct~\cite{llama3}. 

\textbf{Infrastructure.} All fine-tuning experiments were run on a single Standard NC40ads H100 v5 (40 vcpus, 320 GiB memory) GPU on Microsoft Azure. Across all experiments, including model ablations, the total compute usage was 12.54 GPU hours. We used the same machine for data-generation experiments with open-source models. We used Azure's OpenAI API to sample from GPT-5.

\subsection{Fine-tuning}

Due to the large number of experiments and model variants, we used Low-Rank Adaptation (LoRA) \cite{lora} for parameter-efficient fine-tuning. We used the following commonly used parameters: rank $r$=16, scaling factor $\alpha$=32, and a dropout rate of 0.05. LoRA adapters were applied to all linear layers. Models were fine-tuned for two epochs with a batch size of 16, a learning rate of $10^{-4}$, a cosine learning rate scheduler, and the AdamW optimizer. All experiments were conducted using bfloat16 precision (\texttt{bf16=True}). HuggingFace's \cite{hf} \texttt{transformers} and \texttt{peft} libraries were used for fine-tuning.

For sampling from fine-tuned models, we followed common practice and used the following sampling parameters: temperature=0.7, top-p=0.8, top-k=20, and min-p=0.0. Prompt templates are shown in \autoref{fig:ft-prompts}; the same prompt templates were used for formatting the training data. All code snippets in training and sampling data (including skeleton code snippets) are wrapped in \texttt{<code>} and \texttt{</code>}. 

\subsection{Prompting}
For sampling from prompt-based models, we followed common practice and used the following parameters: (1) Qwen2.5-Coder-7B-Instruct, Qwen3-8B, and Llama-3.1-8B-Instruct (temperature=0.7, top-p=0.8, top-k=20, and min-p=0.0), (2) GPT-5 (temperature=1.0, top-p=1.0).
We used the following prompt for all prompted models as shown in \autoref{fig:prompting-prompt}.

\section{Results}
\label{app:results}

\input{latex/figs/combined_app_newq}
\input{latex/tables/embed-app-newq}
\input{latex/figs/embed_app_newq}
\input{latex/tables/llama-func-style}
\input{latex/tables/auc_detail}

This section provides extended results complementing the main paper. We organize the following sections:
\begin{itemize}
    \item \textbf{Additional problem sets.} We report results on \texttt{test\_OP\_2} and \texttt{test\_NP\_2} to evaluate model performance on additional sets of test problems. Functionality and style results are shown in \autoref{fig:combined-app-newq}, embedding-based metrics in \autoref{tab:embedding-metrics-app-newq} and \autoref{fig:embd-app-newq}. Across these additional problems, we observe patterns consistent with \texttt{test\_OP\_1} and \texttt{test\_NP\_1}: \textsc{ParaStudent} variants consistently achieve closer alignment with real student code than prompt-based baselines.
    
    \item \textbf{Detailed engagement AUCs.} \autoref{auc_detail} reports ROC AUCs with 95\% confidence intervals for all models on the aggregated \texttt{test\_NP} split under Experiment~2, using the 25th- and 50th-percentile thresholds of real engagement scores.
    
    
    \item \textbf{Fine-tuning ablation.} We report Wasserstein distance results for the Llama-3.1-8B ablation (\autoref{tab:llama-wasserstein}). The trend echoes that of the Qwen models: the fine-tuned \texttt{llama-parastudent} achieves substantially closer distributional alignment with real student code than its instruction-tuned counterpart \texttt{llama-inst} across all metrics.


\end{itemize}
\input{latex/figs/judge-prompt}
\input{latex/tables/agreement}

\section{LLM Judge for Feedback Engagement}
\label{app:agreement}

\subsection{Prompt}

\autoref{fig:judge-prompt} shows the LLM judge prompt used to label feedback sentence engagement using GPT-4.1~\cite{gpt41}. The prompt provides the problem context, full tutor feedback, a target feedback sentence, successive student code submissions, and a code diff summary, and tasks the judge with determining whether the sentence was relevant to the code edits and whether the resulting changes were correct.

\subsection{Human Validation Instruction to Annotators}

Two annotators reviewed 50 randomly sampled feedback--code edit pairs to validate the labels produced by the LLM judge. For each example, annotators were shown the full problem context, the complete AI tutor feedback, a target feedback sentence, successive student code submissions, and a code difference summary highlighting the changes between submissions. Their task was to assess whether the target feedback sentence was relevant to the student's code edits and, if so, whether the resulting edits were applied correctly.

Specifically, annotators provided the following labels:
\begin{itemize}
    \item \texttt{is\_relevant}: \texttt{true} if the student's code edits were directly motivated by the feedback sentence, and \texttt{false} otherwise.
    \item \texttt{is\_success}: \texttt{true} if the edit fixed or improved the targeted issue; \texttt{false} if the edit was incorrect or introduced new errors; and \texttt{null} if the feedback sentence was not relevant to the edit (\texttt{is\_relevant = false}).
\end{itemize}

Both annotators were authors of the paper and were familiar with the task setting and course content. Annotators were informed that all examples consisted of anonymized student code and AI-generated feedback from an introductory programming course and contained no personally identifiable information. The task involved minimal risk, and annotators could stop at any time. No monetary compensation was provided. The annotation interface consisted of a spreadsheet with predefined fields for the labels.

\subsection{LLM-Human Agreement}

We assessed the reliability of the LLM judge using 50 randomly sampled feedback--code edit pairs annotated independently by two human annotators. Because these labels are highly context-dependent, human annotation is time-consuming, which motivated a targeted validation rather than a large-scale labeling effort. We report pairwise agreement between the LLM judge and each human annotator, as well as human--human agreement, using Cohen's $\kappa$~\cite{cohen1960coefficient}. As shown in \autoref{tab:lm_judge_agreement}, agreement was moderate to substantial across both \textit{Relevant} and \textit{Success} labels.

\section{AI Assistance}
We used AI for writing assistance and code, including but not limited to Codex and Claude Code.

%% file: latex/figs/ft-prompt.tex
\begin{figure*}[t]
\centering

\begin{minipage}[t]{0.48\textwidth}
\centering
\begin{tcolorbox}[
  colback=gray!5,
  colframe=black,
  boxrule=0.5pt,
  arc=2pt,
  width=\textwidth
]
\begin{lstlisting}[
  basicstyle=\ttfamily\scriptsize,
  breaklines=true,
  columns=fullflexible
]
PROBLEM INSTRUCTIONS
{instructions}

FIXED CODE
<code>{fixed_code}</code>

TODO CODE
<code>{skeleton_code}</code>

SUBMISSION HISTORY (CODE)
{submissions_only}
\end{lstlisting}
\end{tcolorbox}
\caption*{(a) Experiment 1}
\label{fig:ft-prompt-exp1}
\end{minipage}
\hfill
\begin{minipage}[t]{0.48\textwidth}
\centering
\begin{tcolorbox}[
  colback=gray!5,
  colframe=black,
  boxrule=0.5pt,
  arc=2pt,
  width=\textwidth
]
\begin{lstlisting}[
  basicstyle=\ttfamily\scriptsize,
  breaklines=true,
  columns=fullflexible
]
PROBLEM INSTRUCTIONS
{instructions}

FIXED CODE
<code>{fixed_code}</code>

TODO CODE
<code>{skeleton_code}</code>

SUBMISSION HISTORY (CODE + FEEDBACK)
{submissions_with_feedback}
\end{lstlisting}
\end{tcolorbox}
\caption*{(b) Experiment 2}
\label{fig:ft-prompt-exp1}
\end{minipage}

\caption{\textbf{Prompt templates for fine-tuned models.} Left (Experiment 1 without feedback), Right (Experiment 2 with feedback).}
\label{fig:ft-prompts}
\end{figure*}

%% file: latex/figs/prompting-prompt.tex
\begin{figure*}[t]
\centering

\lstset{
  basicstyle=\ttfamily\scriptsize,
  breaklines=true,
  columns=fullflexible,
  showstringspaces=false,
  aboveskip=0pt,
  belowskip=0pt
}

\begin{subfigure}[t]{0.48\textwidth}
\vspace{0pt}
\centering
\begin{tcolorbox}[
  colback=gray!5,
  colframe=black,
  boxrule=0.5pt,
  arc=2pt,
  width=\linewidth,
  equal height group=promptingboxes
]
{\footnotesize\bfseries System message}
\par\smallskip
\begin{lstlisting}
YOUR TASK
You are simulating a student taking an introduction to Python programming course.
Generate the student's next attempt at the TODO CODE of the current problem, conditioned on their previous code submissions.

* Only write the contents of the TODO CODE. Do not include docstrings or anything outside it.
* As a novice student, your attempts may include mistakes or partial fixes.
* FIXED CODE is provided; do not copy, modify, or include it.
* Wrap the generated code in <code> and </code> tags so that it can be easily extracted.
* If the generated attempt is the student's final submission, append "<SUBMIT>" at the end.
* Use Python programming language
\end{lstlisting}

\tcblower

{\footnotesize\bfseries User message}
\par\smallskip
\begin{lstlisting}
PROBLEM INSTRUCTIONS
{instructions}

FIXED CODE
<code>{fixed_code}</code>

TODO CODE
<code>{skeleton_code}</code>

SUBMISSION HISTORY (CODE)
{submissions_only}
\end{lstlisting}
\end{tcolorbox}
\caption{Experiment 1}
\label{fig:prompting-exp1}
\end{subfigure}
\hfill
\begin{subfigure}[t]{0.48\textwidth}
\vspace{0pt}
\centering
\begin{tcolorbox}[
  colback=gray!5,
  colframe=black,
  boxrule=0.5pt,
  arc=2pt,
  width=\linewidth,
  equal height group=promptingboxes
]
{\footnotesize\bfseries System message}
\par\smallskip
\begin{lstlisting}
YOUR TASK
You are simulating a student taking an introduction to Python programming course.
Generate the student's next attempt at the TODO CODE of the current problem, using both their prior submissions and the feedback they received.

* Only write the contents of the TODO CODE. Do not include docstrings or anything outside it.
* As a novice student, your attempts may include mistakes or partial fixes.
* FIXED CODE is provided; do not copy, modify, or include it.
* Use prior feedback to inform your code revision.
* Wrap the generated code in <code> and </code> tags so that it can be easily extracted.
* If the generated attempt is the student's final submission, append "<SUBMIT>" at the end.
* Use Python programming language
\end{lstlisting}

\tcblower

{\footnotesize\bfseries User message}
\par\smallskip
\begin{lstlisting}
PROBLEM INSTRUCTIONS
{instructions}

FIXED CODE
<code>{fixed_code}</code>

TODO CODE
<code>{skeleton_code}</code>

SUBMISSION HISTORY (CODE + FEEDBACK)
{submissions_with_feedback}
\end{lstlisting}
\end{tcolorbox}
\caption{Experiment 2}
\label{fig:prompting-exp2}
\end{subfigure}

\caption{\textbf{Prompt templates for prompt-based models.} Left (Experiment 1 without feedback), Right (Experiment 2 with feedback). These prompts are adapted from prior work \cite{student_code_gen2}, which used LLMs to mimic the distribution of error types and test case failure frequencies of real student datasets.}
\label{fig:prompting-prompt}
\end{figure*}

%% file: latex/figs/combined_app_newq.tex
\begin{figure*}[t]
\centering
\begin{subfigure}[t]{0.49\textwidth}
    \centering
    \includegraphics[width=\linewidth]{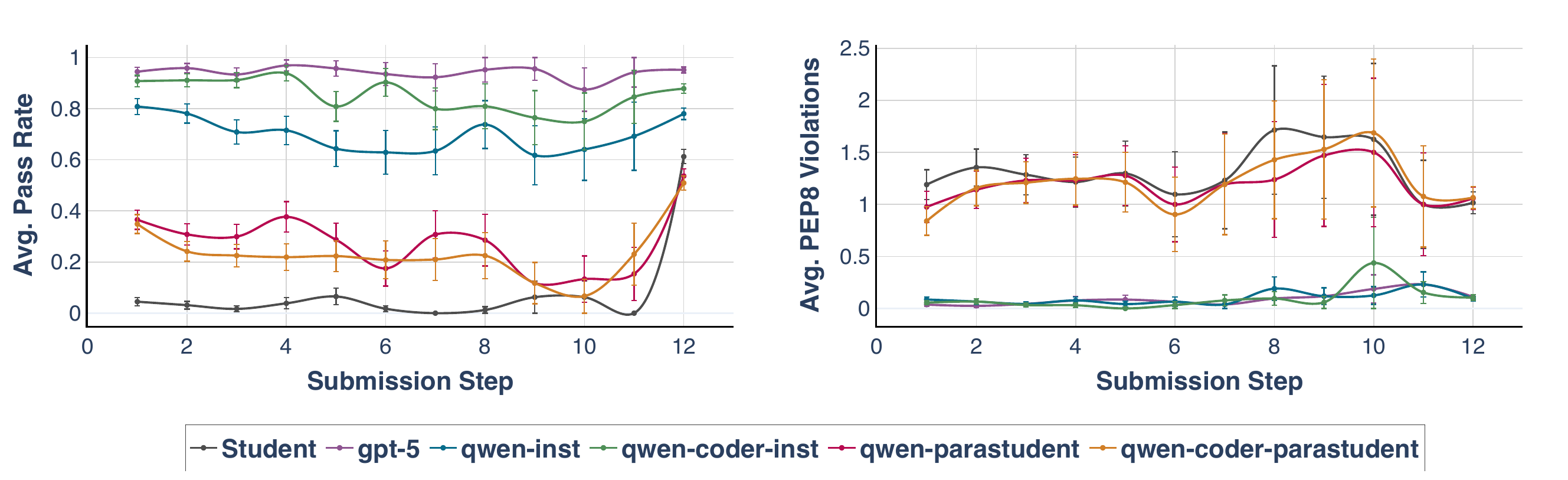}
    \caption{\texttt{test\_OP\_2}}
    \label{fig:combined-app-newq-a}
\end{subfigure}
\hfill
\begin{subfigure}[t]{0.49\textwidth}
    \centering
    \includegraphics[width=\linewidth]{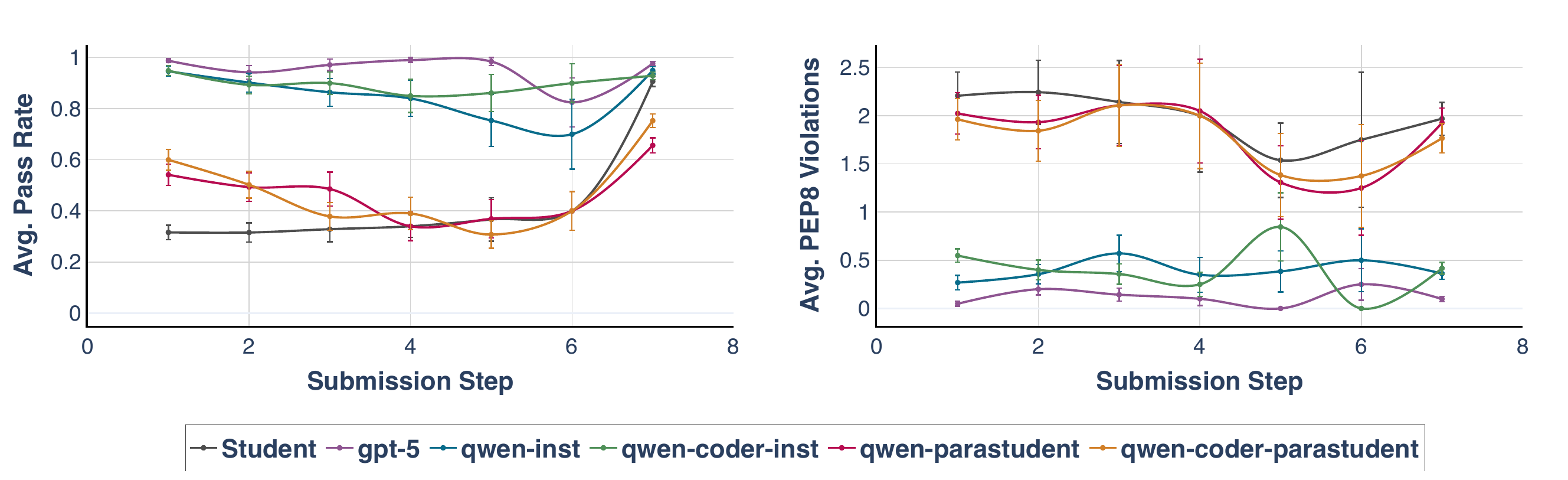}
    \caption{\texttt{test\_NP\_2}}
    \label{fig:combined-app-newq-b}
\end{subfigure}

\caption{\textbf{Progression of pass rate (left) and PEP 8 violation count (right) across submission steps on \texttt{test\_OP\_2} and \texttt{test\_NP\_2} (Experiment 2: with AI tutor feedback).}}
\label{fig:combined-app-newq}
\end{figure*}

%% file: latex/tables/embed-app-newq.tex
\begin{table*}[t]
\caption{\textbf{Embedding-based Similarity Metrics on \texttt{test\_OP\_2} and \texttt{test\_NP\_2}.  (Experiment 1: without AI tutor feedback, Experiment 2: with AI tutor feedback)}}
\label{tab:embedding-metrics-app-newq}
\centering
\scriptsize
\setlength{\tabcolsep}{4pt}
\renewcommand{\arraystretch}{0.85}
\begin{subtable}[t]{0.48\textwidth}
\caption{\texttt{test\_OP\_2}}
\label{tab:embedding-metrics-app-newq-a}
\centering
\begin{tabular}{l|c|c|cc}
\toprule
\textbf{Model} & \textbf{Exp.} & \textbf{Cosine Distance ↓} & \multicolumn{2}{c}{\textbf{Coverage (\%) ↑}} \\
 & & & \textbf{k=5} & \textbf{k=10} \\
\midrule
\texttt{gpt-5}                  & 1 & 0.056 & 40.1 & 53.5 \\
\texttt{qwen-inst}              & 1 & 0.078 & 32.2 & 46.4 \\
\texttt{qwen-coder-inst}        & 1 & 0.082 & 33.2 & 49.5 \\
\texttt{qwen-parastudent}       & 1 & 0.025 & \textbf{94.8} & \textbf{97.8} \\
\texttt{qwen-coder-parastudent} & 1 & \textbf{0.023} & 94.3 & 97.7 \\
\midrule
\texttt{gpt-5}                  & 2 & 0.043 & 59.8 & 73.5 \\
\texttt{qwen-inst}              & 2 & 0.057 & 57.0 & 72.0 \\
\texttt{qwen-coder-inst}        & 2 & 0.064 & 52.9 & 70.6 \\
\texttt{qwen-parastudent}       & 2 & 0.024 & 92.8 & 96.1 \\
\texttt{qwen-coder-parastudent} & 2 & \textbf{0.023} & 93.9 & 97.2 \\
\bottomrule
\end{tabular}
\end{subtable}
\hfill
\begin{subtable}[t]{0.48\textwidth}
\caption{\texttt{test\_NP\_2}}
\label{tab:embedding-metrics-app-newq-b}
\centering
\begin{tabular}{l|c|c|cc}
\toprule
\textbf{Model} & \textbf{Exp.} & \textbf{Cosine Distance ↓} & \multicolumn{2}{c}{\textbf{Coverage (\%) ↑}} \\
 & & & \textbf{k=5} & \textbf{k=10} \\
\midrule
\texttt{gpt-5}                  & 1 & 0.041 & 20.3 & 30.4 \\
\texttt{qwen-inst}              & 1 & 0.025 & 37.3 & 45.4 \\
\texttt{qwen-coder-inst}        & 1 & 0.024 & 57.1 & 72.1 \\
\texttt{qwen-parastudent}       & 1 & 0.016 & 94.7 & 97.8 \\
\texttt{qwen-coder-parastudent} & 1 & 0.015 & \textbf{96.1} & \textbf{98.3} \\
\midrule
\texttt{gpt-5}                  & 2 & 0.034 & 49.0 & 56.0 \\
\texttt{qwen-inst}              & 2 & 0.027 & 52.4 & 62.4 \\
\texttt{qwen-coder-inst}        & 2 & 0.025 & 59.1 & 69.6 \\
\texttt{qwen-parastudent}       & 2 & \textbf{0.014} & 93.6 & 96.4 \\
\texttt{qwen-coder-parastudent} & 2 & 0.015 & 95.5 & 96.9 \\
\bottomrule
\end{tabular}
\end{subtable}
\end{table*}

%% file: latex/figs/embed_app_newq.tex
\begin{figure*}[t]
    \centering
    \begin{subfigure}[t]{0.49\textwidth}
        \centering
        \includegraphics[width=\linewidth]{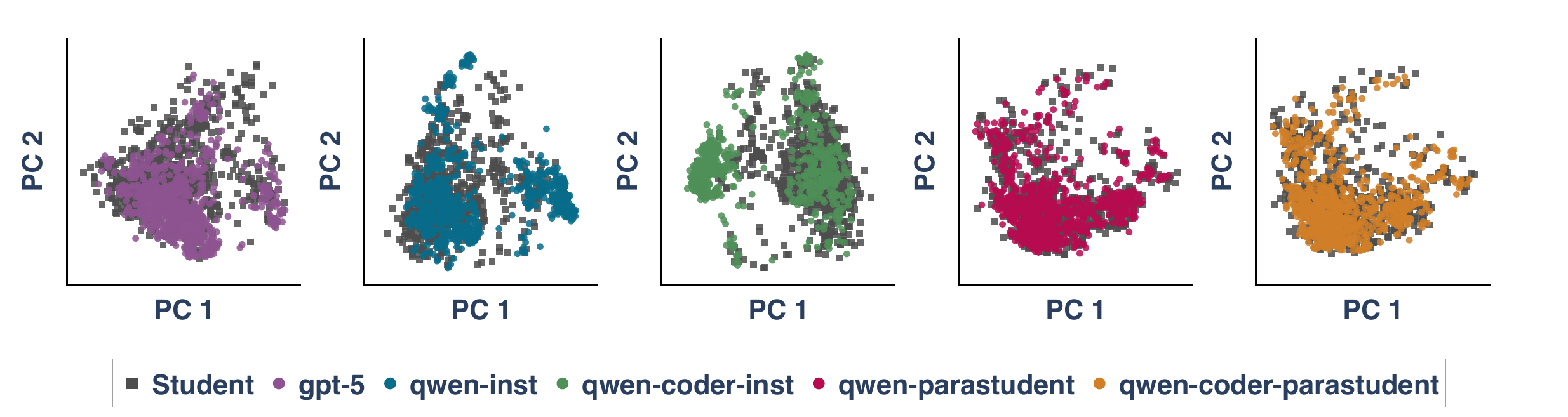}
        \caption{\texttt{test\_OP\_2}}
        \label{fig:embd-app-newq-a}
    \end{subfigure}
    \hfill
    \begin{subfigure}[t]{0.49\textwidth}
        \centering 
        \includegraphics[width=\linewidth]{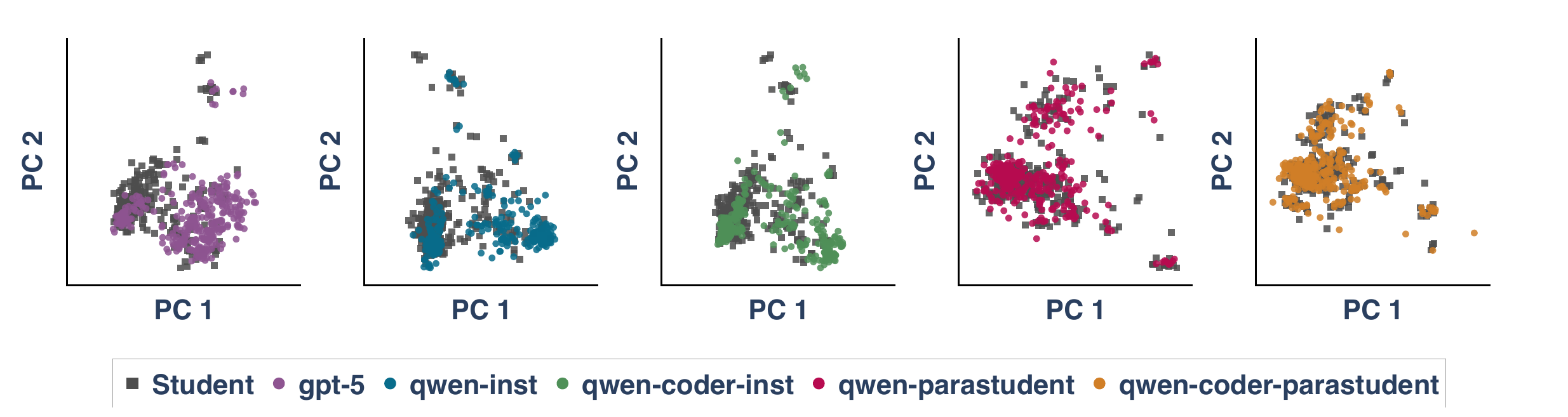}
        \caption{\texttt{test\_NP\_2}}
        \label{fig:embd-app-newq-b}
    \end{subfigure}

\caption{\textbf{Code embedding visualization  on \texttt{test\_OP\_2} and \texttt{test\_NP\_2}. (Experiment 2: with AI tutor feedback)}. Real student code (gray squares) is compared against \texttt{gpt-5} (purple circles), \texttt{qwen-inst} (blue circles), \texttt{qwen-coder-inst} (green circles),  \texttt{qwen-parastudent} (pink circles),  and \texttt{qwen-coder-parastudent} (orange circles).}
    \label{fig:embd-app-newq}
\end{figure*}

%% file: latex/tables/llama-func-style.tex
\begin{table*}[t]
\caption{\textbf{Wasserstein distance between model and student distributions on \texttt{test\_OP\_1} and \texttt{test\_NP\_1} on Llama-3.1-8B model variants. (Experiment
2: with AI tutor feedback)}}
\label{tab:llama-wasserstein}
\centering
\scriptsize
\setlength{\tabcolsep}{4pt}
\renewcommand{\arraystretch}{0.95}
\begin{subtable}[t]{0.49\textwidth}
\centering
\caption{\texttt{test\_OP\_1}}
\label{tab:llama-wasserstein-a}
\resizebox{\textwidth}{!}{%
\begin{tabular}{l|c|c|c|c|c}
\toprule
\textbf{Model} & \textbf{Pass Rate ↓} & \textbf{PEP 8 ↓} & \textbf{LOC ↓} & \textbf{AST Depth ↓} & \textbf{AST Width ↓} \\
\midrule
\texttt{llama-inst}        & 0.138 & 1.033 & 0.801 & 0.511 & 0.697 \\
\texttt{llama-parastudent} & \textbf{0.025} & \textbf{0.231} & \textbf{0.366} & \textbf{0.304} & \textbf{0.539} \\
\bottomrule
\end{tabular}%
}
\end{subtable}
\hfill
\begin{subtable}[t]{0.49\textwidth}
\centering
\caption{\texttt{test\_NP\_1}}
\label{tab:llama-wasserstein-b}
\resizebox{\textwidth}{!}{%
\begin{tabular}{l|c|c|c|c|c}
\toprule
\textbf{Model} & \textbf{Pass Rate ↓} & \textbf{PEP 8 ↓} & \textbf{LOC ↓} & \textbf{AST Depth ↓} & \textbf{AST Width ↓} \\
\midrule
\texttt{llama-inst}        & 0.368 & 0.404 & 1.549 & 0.467 & 0.743 \\
\texttt{llama-parastudent} & \textbf{0.062} & \textbf{0.100} & \textbf{0.546} & \textbf{0.126} & \textbf{0.247} \\
\bottomrule
\end{tabular}%
}
\end{subtable}
\end{table*}

%% file: latex/tables/auc_detail.tex
\begin{table*}[t]
\centering
\small
\setlength{\tabcolsep}{6pt}
\caption{\textbf{Engagement discrimination on the aggregated \texttt{test\_NP} split under Experiment~2 (with AI tutor feedback).} ROC AUCs measure discrimination between interaction streams with real engagement above versus at or below the 25th- or 50th-percentile threshold. Brackets report 95\% confidence intervals. Bold denotes the highest unrounded AUC for each engagement metric and threshold, not statistical significance.}
\begin{tabular}{@{}llcc@{}}
\toprule
\textbf{Score} & \textbf{Model}
& \textbf{AUC: 25th percentile (95\% CI)}
& \textbf{AUC: 50th percentile (95\% CI)} \\
\midrule
$\bar{S}_{\mathrm{rel}}$ & \texttt{gpt-5}
& 0.74 [0.68, 0.80] & 0.77 [0.72, 0.81] \\
$\bar{S}_{\mathrm{rel}}$ & \texttt{qwen-inst}
& 0.75 [0.69, 0.80] & 0.76 [0.71, 0.81] \\
$\bar{S}_{\mathrm{rel}}$ & \texttt{qwen-coder-inst}
& 0.75 [0.69, 0.80] & 0.77 [0.72, 0.82] \\
$\bar{S}_{\mathrm{rel}}$ & \texttt{qwen-parastudent}
& 0.81 [0.76, 0.86] & 0.76 [0.71, 0.81] \\
$\bar{S}_{\mathrm{rel}}$ & \texttt{qwen-coder-parastudent}
& \textbf{0.81} [0.75, 0.87] & \textbf{0.80} [0.75, 0.85] \\
\midrule
$\bar{S}_{\mathrm{succ}}$ & \texttt{gpt-5}
& 0.51 [0.49, 0.55] & 0.52 [0.50, 0.55] \\
$\bar{S}_{\mathrm{succ}}$ & \texttt{qwen-inst}
& 0.57 [0.53, 0.62] & 0.55 [0.51, 0.58] \\
$\bar{S}_{\mathrm{succ}}$ & \texttt{qwen-coder-inst}
& 0.58 [0.53, 0.63] & 0.55 [0.52, 0.59] \\
$\bar{S}_{\mathrm{succ}}$ & \texttt{qwen-parastudent}
& 0.72 [0.67, 0.78] & 0.72 [0.67, 0.77] \\
$\bar{S}_{\mathrm{succ}}$ & \texttt{qwen-coder-parastudent}
& \textbf{0.80} [0.75, 0.85] & \textbf{0.80} [0.75, 0.84] \\
\bottomrule
\end{tabular}
\label{auc_detail}
\end{table*}

%% file: latex/figs/judge-prompt.tex
\begin{figure*}[t]
\centering

\lstset{
  basicstyle=\ttfamily\fontsize{7.5}{7}\selectfont,
  breaklines=true,
  columns=fullflexible,
  keepspaces=true,
  showstringspaces=false,
  aboveskip=0pt,
  belowskip=0pt
}

\begin{tcolorbox}[
  colback=gray!5,
  colframe=black,
  boxrule=0.5pt,
  arc=2pt,
  width=\textwidth,
  boxsep=2pt,
  left=4pt,
  right=4pt,
  top=2pt,
  bottom=2pt
]
{\footnotesize\bfseries System message}
\par\smallskip
\begin{lstlisting}
You are an expert in programming education. These problems come from an introductory programming course where students iteratively revise their code based on feedback. Your task is to analyze **the specified target sentence** in the feedback and explain whether and how it influenced the student's code edits (both additions and deletions), and whether those edits were correct.

You are given:
- The problem statement
- The student's code at time t-1
- The feedback message received at time t-1
- Target sentence to analyze from the feedback
- The student's code at time t
- A summary of the code diff (i.e., the list of lines added and removed between the two versions)

### TASK
1. For the specified sentence to analyze, decide whether the student's code edits (added or removed lines) are **grounded in** (i.e., directly motivated by) that sentence.
2. If the sentence is relevant, judge whether the resulting code edit was correct (i.e., whether the change addressed the feedback successfully) by examining the relevant code edits and inferring the student's reasoning or possible misconception.
3. For the target sentence, first provide your rationale, explaining your reasoning clearly before giving boolean decisions.
4. Always output rationales even when the sentence is irrelevant.

### OUTPUT FORMAT (JSON)
```json
{
  "feedback_analysis": [
    {
      "relevance_rationale": "...",
      "is_relevant": true,
      "success_rationale": "...",
      "is_success": true
    }
  ]
}
```

### NOTES
- **`is_relevant` should be:**
 - `true` -> the student's code edits were directly motivated by this feedback sentence
 - `false` -> the student's code edits were not influenced by this sentence
- **`is_success` should be:**
 - `true` -> the edit fixed or improved the targeted issue
 - `false` -> the edit was incorrect or introduced new errors
 - `null` -> the sentence was irrelevant to the code edit (`is_relevant = false`)
- When generating rationales:
 - For **`relevance_rationale`**, focus on **feedback--edit causality**, referring to specific added/removed lines.
 - For **`success_rationale`**, focus on **whether the edit successfully addresses the targeted feedback** and inferred student understanding or misconception.
 - Place each rationale immediately before its corresponding boolean field in the JSON.
- Output **JSON only**, with **no commentary** outside the JSON.

\end{lstlisting}

\tcblower

{\footnotesize\bfseries User message}
\par\smallskip
\begin{lstlisting}
PROBLEM INSTRUCTIONS:
{instructions}

FULL FEEDBACK:
{feedback_text}

TARGET SENTENCE TO ANALYZE:
{sentence}

BLOCK TRANSITION:
Block t-1 -> t ({label})

CODE AT TIME t-1:
<code>
{prev_code}
</code>

CODE AT TIME t:
<code>
{next_code}
</code>

CODE DIFF SUMMARY:
ADDED:
{added_line}

REMOVED:
{removed_line}
\end{lstlisting}
\end{tcolorbox}

\caption{\textbf{Prompt used for the LLM judge to attribute student code edits to individual feedback sentences.}}
\label{fig:judge-prompt}
\end{figure*}

%% file: latex/tables/agreement.tex
\begin{table*}[!t]
\caption{\textbf{Pairwise agreement between the LLM judge and two human annotators (A1 and A2), and between annotators, for \textit{Relevant} and \textit{Success} labels on 50 randomly sampled feedback--code edit pairs. Positive and negative agreement are computed with respect to the human annotator for LLM--human comparisons and A1 for the A1--A2 comparison.}}
\label{tab:lm_judge_agreement}
\centering
\small
\setlength{\tabcolsep}{6pt}
\begin{tabular}{@{}lcccccccc@{}}
\toprule
& \multicolumn{4}{c}{\textbf{Relevant}}
& \multicolumn{4}{c}{\textbf{Success}} \\
\cmidrule(lr){2-5}\cmidrule(lr){6-9}
\textbf{Comparison Pair}
& \textbf{Positive} & \textbf{Negative} & \textbf{All}
& $\boldsymbol{\kappa}$
& \textbf{Positive} & \textbf{Negative} & \textbf{All}
& $\boldsymbol{\kappa}$ \\
\midrule
LLM--A1 & 0.96 & 0.64 & 0.80 & 0.60
        & 0.68 & 0.97 & 0.86 & 0.69 \\
LLM--A2 & 0.93 & 0.71 & 0.84 & 0.66
        & 0.60 & 0.93 & 0.80 & 0.56 \\
A1--A2  & 0.96 & 0.80 & 0.88 & 0.76
        & 0.89 & 0.90 & 0.90 & 0.79 \\
\bottomrule
\end{tabular}
\end{table*}